\let\oldyear\year
\documentclass{ieeeaccess}

\let\year\oldyear

\usepackage{graphicx}
\usepackage{amsmath}
\usepackage{amssymb}
\usepackage{amsfonts}
\usepackage{algorithm}
\usepackage{algpseudocode}
\usepackage{booktabs}
\usepackage{cite}
\usepackage{url}
\usepackage{microtype}
\usepackage{float}
\usepackage{tabularx}
\usepackage{listings}
\usepackage{algorithm}
\usepackage{algpseudocode}
\usepackage{caption}
\usepackage{textcomp}

\definecolor{codegray}{rgb}{0.5,0.5,0.5}
\definecolor{codeblue}{rgb}{0.0,0.3,0.7}
\definecolor{codegreen}{rgb}{0.1,0.5,0.1}

\begin{document}
\history{Date of publication xxxx 00, 2026, date of current version xxxx 00, 2026.}
\doi{10.1109/ACCESS.2026.DOI}

\title{Federated Stream-Processing and Latency-Gated Response for Cross-Sector Threat Detection and Collaborative Containment}

\author{\uppercase{Namit Mohale}}
\address{Independent Researcher, San Francisco, CA, USA}

\tfootnote{This research received no specific grant from any funding agency in the public, commercial, or not-for-profit sectors.}
\corresp{Corresponding author: Namit Mohale (e-mail: mohale.namit95@gmail.com).}

\begin{abstract}
Critical infrastructure defense is fundamentally bottlenecked by the operational reality that preventive controls are frequently bypassed by sophisticated supply-chain compromises and stolen administrative credentials. When prevention fails, defense relies entirely on rapid, post-ingress threat detection and automated response across sovereign sectors. We present a novel, federated, high-throughput stream-processing and correlation framework designed to detect coordinated cross-sector threat campaigns and orchestrate containment at machine speed. By utilizing a stateless Pre-Filtering Dispatcher Subsystem (PFDS), in-memory lock-sharded state workers, and a $95\%$ statistical watermark heuristic, our system maintains detection momentum during network partitions to evaluate speculative alerts. Delayed telemetry is subsequently reconciled directly within a version-keyed columnar storage engine via deterministic time-bucket hashing, eliminating state-retraction overhead. We evaluate a prototype of our framework—implemented in Go with an instantiated production-grade columnar analytical store—against a $500,000$ events per second workload. The results demonstrate an internal framework processing overhead of under \textbf{7 seconds}, while achieving total end-to-end operational convergence-accounting for multi-sector detection, correlation, wide-area network (WAN) propagation, windowing stability, VLAN-level response, and hardware level mitigation commitment-within a realistic \textbf{12-20 seconds} window.
\end{abstract}


\titlepgskip = -15pt

\maketitle

\section{Introduction}
\label{sec:intro}

\PARstart{C}{ritical} infrastructure sectors—such as healthcare systems, energy grids, water treatment facilities, and transportation networks—have become the primary targets of highly sophisticated, state-sponsored Advanced Persistent Threat (APT) actors. Unlike conventional enterprise IT environments, intrusions into operational technology (OT) and industrial control systems (ICS) carry physical-world consequences, ranging from power grid blackouts to the compromise of life-support systems in regional hospitals. 

Historically, defense paradigms have focused heavily on perimeter-based \textit{prevention}, employing firewalls, strict network segregation, and identity access controls. However, modern operational realities have exposed a fatal flaw in this preventive-first strategy: **prevention is inherently bypassed when adversaries leverage legitimate access vectors.** By compromising shared supply-chain elements, external maintenance service providers, or common third-party identity credentials, an adversary can slip past perimeters undetected. Because these independent critical sectors are physically and logically isolated from one another, and because privacy regulations (such as HIPAA in healthcare or NERC CIP in utilities) strictly prohibit them from sharing raw telemetry, an adversary can execute a coordinated, cross-sector campaign completely in parallel. To each isolated security operations center (SOC), the attack appears as a minor, localized, and uncorrelated anomaly, leaving the broader campaign entirely invisible until catastrophic physical damage occurs.

Consequently, modern critical infrastructure defense must undergo a paradigm shift: **assuming breach and prioritizing rapid, post-ingress detection and automated response (containment).** Rather than attempting the impossible task of preventing every credential abuse or supply-chain compromise, defenders must focus on minimizing the ``blast radius'' once an intrusion has occurred. 

To outpace active adversaries and limit lateral damage across isolated sectors, a collaborative threat detection and response framework must operate under three uncompromising constraints:
\begin{itemize}
    \item \textbf{Machine-Speed Detection and Containment:} Post-ingress detection and subsequent containment actions must execute within a strict $20$-second latency budget from the initial edge compromise to prevent physical damage.
    \item \textbf{Late-Data Resilience:} The detection engine must remain highly resilient to out-of-order, delayed telemetry streams (``stragglers'') caused by degraded physical links, satellite bottlenecks, or edge reboots, without stalling detection in healthy sectors.
    \item \textbf{Sovereign Privacy Preservation:} The system must achieve deep cross-sector threat correlation and coordinate response actions without ever transmitting raw event logs across sector boundaries.
\end{itemize}

\subsection{Limitations of the State-of-the-Art}
Existing security toolsets are structurally incapable of satisfying these detection and response constraints. Traditional Security Information and Event Management (SIEM) systems and centralized security data lakes are fundamentally too slow; they rely on polling-based query models that introduce minutes of delay, rendering real-time containment impossible. 

To bypass this latency bottleneck, general-purpose Distributed Stream Processing Engines (DSPEs) like Apache Flink have been explored. However, these engines enforce a strict, binary Event-Time consistency model. If a single edge network partition stalls, the global detection watermark freezes. This ``straggler problem'' stalls the entire cross-sector detection pipeline, blinding defenders to active threats in healthy sectors. Furthermore, when these engines attempt to reconcile delayed logs for forensic response, they rely on heavy state backends that cause severe memory thrashing and state starvation under heavy security workloads.

\subsection{Our Proposed Solution: Detection and Reactive Response}
In this paper, we present a novel, federated, high-throughput stream-processing and correlation framework designed specifically for real-time, cross-sector critical infrastructure threat detection and automated containment. 

Our framework abandons the slow, centralized indexing paradigm of traditional SIEMs and the fragile, strictly consistent watermarking models of traditional DSPEs. Instead, we propose a decentralized, \textbf{Rules-as-Code (RaC)} execution model where security engineers write highly expressive, procedurally defined detection graphs via a unified Fluent API. 

The system operates through a highly optimized, multi-tiered pipeline:
\begin{enumerate}
    \item \textbf{The Pre-Filtering Dispatcher Subsystem (PFDS):} This stateless, ultra-low-latency layer sits directly at the ingestion edge. By compiling the stateless, initial bootstrap predicates of our rules into highly optimized Common Expression Language (CEL) ASTs or WebAssembly (Wasm) bytecode, the PFDS filters and discards up to $99.8\%$ of background noise in a single pass to accelerate detection times.
    \item \textbf{Lock-Sharded Stateful Workers:} The filtered, low-volume streams are consumed by concurrent Go-based stream workers. By utilizing a shared-nothing, in-memory partition mapping with key-specific read-write locks, our workers eliminate global mutex contention, allowing parallel, sub-millisecond evaluation of complex sliding temporal windows.
    \item \textbf{95\% Statistical Watermarking \& Speculative Detection:} To handle the inevitable ``straggler problem,'' our watermark coordinator calculates log completeness based on the 95th percentile ($\mathcal{P}_{95}$) of historical latency. If data from a lagging sector falls into the remaining $5\%$, the engine does not stall; it evaluates the active window speculatively and publishes a speculative alert to initiate early-stage containment. 
    \item \textbf{Idempotent Late-Data Reconciliation:} When delayed logs eventually arrive, the engine leverages deterministic time-bucket hashing to generate an identical, unique Fact ID ($F_{\text{id}}$). It writes the revised state back to a highly scalable, columnar analytical database engine designed to natively execute background, version-keyed row replacements. The storage layer automatically overwrites the speculative record with zero manual database administration overhead, and a lightweight revision event is published downstream to update active response measures.
    \item \textbf{Decoupled, Zero-Leak Automated Response:} We implement a secure \textbf{Claim-Check Pattern} over a central CISA-managed Pub/Sub broker. Cooperating sectors never transmit raw logs; instead, they write heavy, sensitive forensic payloads locally and broadcast only lightweight, anonymized Cluster IDs. Downstream, decoupled reactive agents—including an \textbf{Automated Containment Agent} (which evaluates risk gates and signs Ed25519 commands), an advanced \textbf{Gen-AI Triage Agent}, and an \textbf{Escalation Agent}—subscribe to this broker, dynamically reclaiming authorized facts on-demand from the columnar storage layer to execute containment actions (such as local VLAN micro-segmentation or token revocation) without exposing raw logs.
\end{enumerate}

\subsection{Summary of Contributions}
The principal contributions of this work are summarized as follows:
\begin{itemize}
    \item \textbf{The Rules-as-Code Fluent API \& PFDS Compiler:} We present a functional, testable execution language for multi-sector security telemetry, with a compiler capable of separating and pushing down stateless predicates into a high-performance Wasm/CEL Pre-Filtering Dispatcher.
    \item \textbf{Statistical Watermarking \& Memory-Efficient Reconciliation:} We formulate a $95\%$ statistical watermark heuristic that prevents pipeline stalls during network outages, coupled with a deterministic hashing model that offloads late-data state reconciliation directly to the database storage layer.
    \item \textbf{Privacy-Preserving Claim-Check Response Plane:} We design a decoupled, fan-out mitigation architecture that protects sector sovereignty, executing cryptographically signed, risk-gated network and system containment without exposing raw logs outside of local boundaries.
    \item \textbf{Empirical Validation on a National-Scale Testbed:} We implement and benchmark our architecture using Go and ClickHouse-based prototype, subjecting it to a simulated $500,000$ eps workload mimicking a coordinated, multi-vector campaign targeting state, healthcare and utility networks. We demonstrate that the framework introduces only \textbf{sub-7 seconds} of internal computational latency, successfully decoupling software overhead from the unavoidable network and hardware convergence delays. This enables cross-sector correlation and micro-segmentation in under 15 seconds in general scenarios and 20 seconds for more complex scenarios, meeting the critical operational threshold for post-ingress threat containment.
\end{itemize}

\section{Background and Problem Formulation}
\label{sec:background}

In this section, we construct the formal mathematical models for multi-sector event streams, define the mechanics of the late-data ``straggler'' bottleneck in stream-processing architectures, and formulate the core security and privacy constraints of collaborative, cross-sector threat detection.

\subsection{Multi-Sector Stream and Telemetry Model}
Let $\mathcal{S} = \{S_1, S_2, \dots, S_N\}$ be a finite set of $N$ heterogeneous, sovereign critical infrastructure sectors. Each sector $S_i \in \mathcal{S}$ operates an independent network domain and generates an infinite, continuous stream of security, system, and operational telemetry:
\begin{equation}
    E_i = \{e_1, e_2, \dots\}
\end{equation}
Because these sectors are completely network-isolated and air-gapped from one another, their raw telemetry streams are strictly disjoint:
\begin{equation}
    E_i \cap E_j = \emptyset \quad \forall i \neq j
\end{equation}

To allow unified computation over these disparate streams without violating sectoral schema integrity, we model each log record as a standardized tuple within our \textit{Unified Telemetry Schema} (UTS):
\begin{equation}
    e = \langle t_e, t_a, k, \mathbf{x} \rangle
\end{equation}
where:
\begin{itemize}
    \item $t_e \in \mathbb{R}^+$ is the \textit{Event-Time}, representing the physical timestamp when the state transition or security event occurred at the edge asset.
    \item $t_a \in \mathbb{R}^+$ is the \textit{Arrival-Time}, representing the timestamp when the ingestion engine successfully received and normalized the event.
    \item $k \in \mathcal{K}$ is a deterministic, cryptographically hashed partition key (e.g., a one-way SHA-256 hash of a username, source IP, or device MAC address) used to map identity across sectors without exposing raw, sensitive identifiers.
    \item $\mathbf{x} \in \mathcal{X}$ is a polymorphic payload vector containing domain-specific metrics, system calls, or SCADA register parameters.
\end{itemize}

\begin{figure*}[!t]
    \centering
    \includegraphics[width=0.95\textwidth, height=12cm, keepaspectratio]{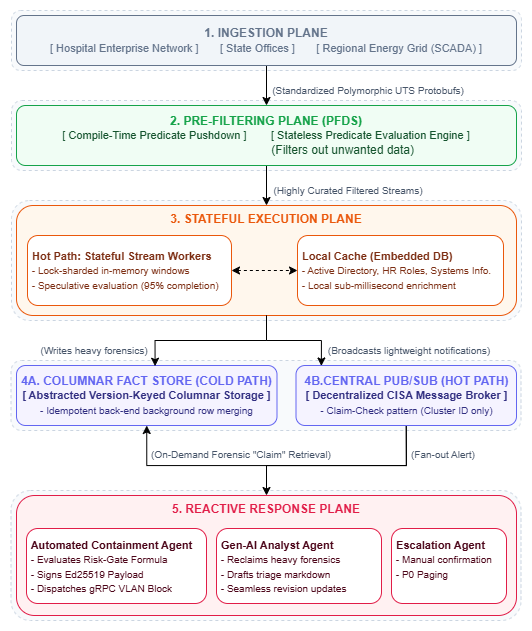}
    \caption{Proposed Federated System Architecture: End-to-end post-ingress threat detection, stateful stream evaluation, and late-data speculative response across independent, sovereign critical sectors.}
    \label{fig:proposed_architecture}
\end{figure*}

\subsection{The Late-Data Straggler Problem}
In any distributed critical infrastructure network, physical propagation delays, WAN congestion, edge node reboots, and satellite link drops are inevitable. We define the propagation delay of an event $e$ as:
\begin{equation}
    \Delta_p(e) = t_a - t_e
\end{equation}

In a perfectly synchronous system, $\Delta_p(e) \approx 0$. However, in real-world critical infrastructure deployments, propagation delays are highly variable and unpredictable.

Let $W(t)$ represent the global system watermark at wall-clock time $t$. Under a \textit{Strictly Consistent Event-Time Model}, a stream-processing engine cannot safely close a temporal processing window of width $\tau$ until it is certain that all data within that window has arrived. Thus, the system's watermark is strictly bounded by the slowest streaming sector:
\begin{equation}
    W(t) = \min_{S_i \in \mathcal{S}} \left( \max_{e \in E_i, t_a \le t}(t_e) \right)
\end{equation}

If a single sector $S_{\text{straggler}} \in \mathcal{S}$ experiences a localized network partition or satellite drop at wall-clock time $t_{\text{outage}}$, its ingestion stream halts. Consequently, the maximum event-time observed from $S_{\text{straggler}}$ stalls at some historical timestamp $t_{\text{stall}}$. Because the global watermark $W(t)$ is constrained by a strict minimum function, it remains frozen at $t_{\text{stall}}$ for the duration of the outage:
\begin{equation}
    W(t) = t_{\text{stall}} \quad \forall t \ge t_{\text{outage}}
\end{equation}

This freeze creates a linear, unbounded growth in the \textit{Systemic Detection Lag} ($L_{\text{sys}}$) across the entire multi-sector pipeline:
\begin{equation}
    L_{\text{sys}}(t) = t - W(t)
\end{equation}

\begin{figure}[H]
    \centering
    \includegraphics[width=\columnwidth]{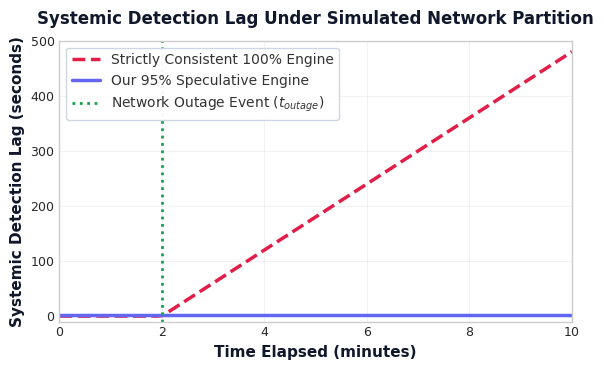}
    \caption{Systemic detection lag ($L_{\text{sys}}$) over time under strict consistency (dashed red line) showing linear growth post-partition, contrasted against the stable, flat baseline of our 95\% speculative model (solid blue line).}
    \label{fig:systemic_lag_concept}
\end{figure}

As illustrated in Figure~\ref{fig:systemic_lag_concept}, this structural vulnerability is highly exploitable. Under standard strictly consistent paradigms, a single localized network drop at $t_{\text{outage}} = 2$ minutes causes $L_{\text{sys}}$ to escalate aggressively over time. By freezing the global watermark, the entire cross-sector detection pipeline becomes blinded, preventing active security monitoring in completely healthy, unaffected sectors. 

In contrast, our proposed $95\%$ statistical watermarking model—formulated mathematically in Section~\ref{sec:architecture} and visualized as a flat baseline in Figure~\ref{fig:systemic_lag_concept}—maintains detection continuity. By bypassing the lagging $5\%$ tail of late-arriving telemetry, the watermark monotonically advances, keeping systemic detection lag flat and bounded below $2$ seconds regardless of ongoing edge network partitions.

Let $\mathcal{F}(S_k(t))$ be a stateful threat detection function operating over an in-memory sliding window $S_k(t)$ of width $\tau$.

\subsubsection{Problem 1 (Watermark Boundary Optimization)}
Our objective is to design a watermark coordinator that dynamically advances $W(t)$ to minimize systemic detection lag $L_{\text{sys}}(t)$ while maintaining a detection confidence bound of $1 - \alpha$ (where $\alpha = 0.05$, representing a 95\% statistical data-completeness guarantee):
\begin{equation}
\begin{aligned}
    \min \quad & \mathbb{E}[L_{\text{sys}}(t)] \\
    \text{s.t.} \quad & \Pr\left( t_a \le W(t) \right) \ge 0.95 \quad \forall e \in E_i
\end{aligned}
\end{equation}

\subsection{Cross-Sector Collaborative Correlation and Response Constraints}
To detect a sophisticated APT executing a synchronized, multi-vector campaign, our framework must correlate events across independent sectors. 

Let a \textit{Coordinated Cross-Sector Campaign} $C$ be defined as a set of localized threat anomalies occurring across $M$ distinct sectors ($M \ge 2$) within a maximum temporal correlation window $\Delta_C$:
\begin{equation}
    C = \{a_{i_1}, a_{i_2}, \dots, a_{i_M}\} \quad \text{where} \quad a_{i_j} \in S_{i_j}
\end{equation}
\begin{equation}
    \max(t_{a_{i_j}}) - \min(t_{a_{i_j}}) \le \Delta_C
\end{equation}
Each localized anomaly $a$ is characterized by a shared, cryptographically hashed identity attribute $k_{\text{actor}} \in \mathcal{K}$.

\subsubsection{Sovereignty and Privacy Constraint}
To comply with zero-trust directives, sector sovereignty, and privacy regulations, we enforce a zero raw-log leakage boundary. Let $\mathcal{L}(e)$ represent the information leakage function of raw telemetry. Our primary privacy constraint dictates that no raw event payload $\mathbf{x}$ may ever be transmitted outside of its originating sector's sovereign network boundary:
\begin{equation}
    \mathcal{L}_{\text{shared}}(e) = 0 \quad \forall e \in E_i \quad \forall S_i \in \mathcal{S}
\end{equation}

\subsubsection{Problem 2 (Zero-Leak Detection and Response Correlation)}
We must define a global correlation and response function $\mathcal{C}$ that operates entirely over a decentralized messaging infrastructure such that:
\begin{equation}
    \mathcal{C}(E_1, E_2, \dots, E_N) \rightarrow \{C_1, C_2, \dots\}
\end{equation}
\begin{equation}
    \text{subject to} \quad \mathcal{L}_{\text{shared}}(e) = 0 \quad \forall e \in E_i
\end{equation}
Our system must achieve this by executing a \textit{Claim-Check Pattern}—writing heavy, localized forensic payloads locally to a secure, columnar database and broadcasting only completely anonymized, lightweight trigger hashes to coordinate automated containment.

\section{Proposed System Architecture}
\label{sec:architecture}

To address the latency, resilience, and privacy constraints formulated in Section \ref{sec:background}, we present the design of our federated threat detection and reactive containment framework. As illustrated in Figure \ref{fig:proposed_architecture}, the architecture is divided into five distinct operational planes: the Ingestion Plane, the Pre-Filtering Dispatcher Subsystem (PFDS), the Stateful Execution Plane, the Storage \& Broker Plane, and the Reactive Response Plane.

\paragraph{Architectural Scope and Boundaries}
While our functional prototype is fully realized using the technologies detailed in Section~\ref{sec:evaluation}, the granular, code-level execution of individual threat-detection rules, local state-machine data transformations, and low-level parsing routines is considered out of scope for this paper. This work focuses strictly on the structural stream-routing mechanics, cryptographic data isolation (via the Claim-Check pattern), and the temporal synchronization guarantees of the architecture. The internal, operational implementation of specific telemetry parsing logic is treated as a modular runtime detail, fully decoupled from the core systemic security and scalability contributions of this framework.

\subsection{Pre-Filtering Dispatcher Subsystem (PFDS)}
The first line of post-ingress defense is the Pre-Filtering Dispatcher Subsystem (PFDS), represented as the second operational tier in our architecture (Plane 2 of Figure~\ref{fig:proposed_architecture}). Because evaluating complex stateful rules over sliding windows is computationally expensive, we deploy stateless, single-pass filtering agents directly at each sector's ingestion edge.

The PFDS evaluates incoming telemetry $E_i$ before serialization or transmission. To achieve sub-millisecond execution, security engineers write stateless bootstrap predicates in a unified, declarative grammar. The PFDS compiler parses these rules and generates Abstract Syntax Trees (ASTs) executed via the \textit{Common Expression Language} (CEL) or compiles them directly into \textit{WebAssembly} (Wasm) bytecode.

For example, if a stateful detection rule seeks to identify brute-force SCADA write attempts, the PFDS compiles a stateless predicate to discard all non-write operations (e.g., standard read requests or system heartbeats) in a single pass. By dropping up to $99.8\%$ of benign background noise at the edge in best-case scenarios, the PFDS ensures that only highly curated, threat-relevant streams are forwarded to the stateful workers, protecting downstream memory and processing queues from exhaustion attacks.

The declarative, developer-friendly syntax of our Rules-as-Code (RaC) engine is illustrated in Listing~\ref{lst:rule_definition}. This fluent specification defines a multi-sector correlation rule that targets brute-force authentication anomalies followed by lateral SCADA register modifications within a sliding 15-minute window.

\begin{lstlisting}[language=Go, caption={Fluent API Rules-as-Code Specification for Cross-Sector Anomalies}, label={lst:rule_definition}]
// Define a cross-sector multi-vector threat rule
var BruteForceToSCADA = NewRule("Correlation-BruteForce-SCADA").
    WithWindow(15 * time.Minute).
    WithKeys("user_hash", "source_ip_hash").
    // First Event: Brute force detection on AD/Identity Sector
    OnStream("identity_events").
    Where(func(e Event) bool {
        return e.GetString("action") == "AUTH_FAIL" && e.GetInt("attempt_count") > 5
    }).
    // Second Event: Modbus SCADA write operation
    OnStream("utility_scada").
    Where(func(e Event) bool {
        return e.GetString("protocol") == "modbus" && e.GetBool("is_write") == true
    }).
    // Correlation logic executed upon match
    Correlate(func(s1 State, s2 State) bool {
        return s1.GetTimestamp("last_fail") < s2.GetTimestamp("write_time")
    })
\end{lstlisting}

To minimize CPU overhead and guarantee deterministic execution bounds at the resource-constrained ingress edge, the PFDS compiles declarative filtering rules into stateless Common Expression Language (CEL) expressions. These expressions are evaluated within a highly optimized, sandboxed WebAssembly (Wasm) runtime embedded directly within the ingestion daemon, completely bypassing heavy language runtime reflection or JSON parsing overhead. By offloading this high-throughput, first-pass filter to a lightweight, compiled execution layer, the subsystem effectively drops the bulk of background telemetry noise with sub-millisecond local processing latency before any WAN transmission occurs.

\subsection{Lock-Sharded Stateful Stream Workers}
The curated stream emitted by the PFDS is consumed by concurrent, stateful stream workers operating within the Stateful Execution Plane (Plane 3 of Figure~\ref{fig:proposed_architecture}). To prevent global mutex contention under high-throughput workloads (such as $500,000$ events per second), we implement a shared-nothing, lock-sharded memory mapping model.

Instead of maintaining a single, globally locked state store, the active memory space is divided into discrete, independent partitions. When an event $e = \langle t_e, t_a, k, \mathbf{x} \rangle$ arrives at a worker, the system hashes the partition key $k$ to allocate a specific worker partition. A localized read-write lock is acquired for that key hash:
\begin{equation}
    \text{ShardID} = \text{MurmurHash3}(k) \pmod{N_{\text{shards}}}
\end{equation}
By sharding locks at the key level rather than the global level, multiple threads can concurrently read and write to independent state slices $S_{k,b}$ without blocking each other. This keeps our lock contention overhead flat and near-zero, even under extreme telemetry surges.

To accelerate contextual lookups during evaluation, each stream worker maintains a highly optimized local cache using an embedded key-value store (instantiated via RocksDB in our prototype). This cache stores localized enrichment context—such as active network topologies, operational system states, and asset-to-owner maps—allowing the engine to enrich incoming telemetry with local context in sub-millisecond intervals without querying external databases.

\subsection{Speculative Detection and 95\% Statistical Watermarking}
To maintain continuous detection capability during network partitions, the framework employs a speculative execution paradigm controlled by a dynamic three-tiered watermarking coordinator.

Instead of tracking watermark progress strictly based on a binary $100\%$ completeness model, the control plane calculates the 95th percentile ($\mathcal{P}_{95}$) of historical transit latency across all active ingestion streams. The watermark $W(t)$ is allowed to monotonically advance even if a lagging sector ($S_{\text{straggler}}$) fails to emit telemetry.

When $W(t)$ advances past a sliding window's end boundary, the stateful worker does not stall. It evaluates the compiled Rule Logic ($\mathcal{F}$) speculatively based on the $95\%$ complete data currently available in memory. If a threshold is crossed, the engine generates a speculative alert, marked with a unique, deterministically generated Fact ID ($F_{\text{id}}$):
\begin{equation}
    F_{\text{id}} = \text{Hash}(\text{RuleID} \parallel k \parallel b)
\end{equation}
By publishing this speculative alert downstream, the system initiates early-stage containment actions before the attacker can propagate laterally, effectively decoupling the speed of incident response from WAN latency or edge network drops.

To formalize the runtime behavior of our stream workers when encountering network partitions, we outline the speculative evaluation heuristic in Algorithm~\ref{alg:speculative_evaluation}.

\begin{algorithm}[H]
\caption{Speculative Window Evaluation}
\label{alg:speculative_evaluation}
\footnotesize
\begin{algorithmic}[1]
\Require State $S_{k,b}$, threshold $\theta = 0.95$, watermark $W(t)$
\Ensure Speculative alert or state reconciliation
\Statex
\State $t_{\text{end}} \gets \text{GetWindowEnd}(b)$
\State $C \gets \text{CalcCompleteness}(S_{k,b}, W(t))$ 
\Statex \hfill \Comment{Ratio of active reporting sectors}

\If{$W(t) \ge t_{\text{end}}$}
    \If{$C \ge \theta$ \textbf{and} $\neg \text{IsFinalized}(S_{k,b})$}
        \State $\text{Alert}_{\text{spec}} \gets \text{EvaluateRule}(\mathcal{F}, S_{k,b})$
        \If{$\text{Alert}_{\text{spec}} \ne \emptyset$}
            \State $F_{\text{id}} \gets \text{GenFactID}(\text{RuleID}, k, b)$
            \State $\text{EmitAlert}(\text{Alert}_{\text{spec}}, F_{\text{id}}, \text{"Speculative"})$
            \State $\text{SetSpeculative}(S_{k,b}, \text{true})$
        \EndIf
    \EndIf
\EndIf

\Statex
\Procedure{ReconcileLateData}{Event $e_{\text{late}}$}
    \State $b \gets \text{GetTimeBucket}(e_{\text{late}}.t_e)$
    \State $\text{UpdateWindowState}(S_{k,b}, e_{\text{late}})$
    \If{$\text{IsSpeculative}(S_{k,b})$}
        \State $\text{Alert}_{\text{final}} \gets \text{EvaluateRule}(\mathcal{F}, S_{k,b})$
        \State $F_{\text{id}} \gets \text{GenFactID}(\text{RuleID}, k, b)$
        \State $\text{EmitAlert}(\text{Alert}_{\text{final}}, F_{\text{id}}, \text{"Finalized"})$
        \State $\text{SetFinalized}(S_{k,b}, \text{true})$
    \EndIf
\EndProcedure
\end{algorithmic}
\end{algorithm}

Complementary to window evaluation, the control plane must dynamically adjust the system watermark based on real-time ingestion latency. This process is formalized in Algorithm~\ref{alg:watermark_calibration}.

\begin{algorithm}[H]
\caption{Dynamic Watermark Calibration}
\label{alg:watermark_calibration}
\footnotesize
\begin{algorithmic}[1]
\Require Sectors $\mathcal{S}$, buffer $H$ (size $M$), percentile $\mathcal{P}_{95}$
\Ensure System watermark $W(t)$
\Statex
\For{each ingested event $e = \langle t_e, t_a, k, \mathbf{x} \rangle$ from $S_i$}
    \State $\Delta_p(e) \gets t_a - t_e$ \Comment{Event propagation delay}
    \State $\text{PushToBuffer}(H_i, \Delta_p(e))$
    \If{$\text{BufferSize}(H_i) > M$}
        \State $\text{PopOldest}(H_i)$
    \EndIf
\EndFor

\Statex
\Procedure{OnWatermarkIntervalTick}{}
    \State $\mathbf{L} \gets \emptyset$
    \For{each sector $S_i \in \mathcal{S}$}
        \State $L_i \gets \text{CalcPercentile}(H_i, \mathcal{P}_{95})$
        \State $\mathbf{L} \gets \mathbf{L} \cup \{L_i\}$
    \EndFor
    
    \State $L_{\text{sys\_bound}} \gets \text{Mean}(\mathbf{L})$
    \State $t_{\text{current\_wall}} \gets \text{GetWallClockTime}()$
    
    \State $W(t) \gets t_{\text{current\_wall}} - L_{\text{sys\_bound}}$
    \State $\text{BroadcastWatermark}(W(t))$
\EndProcedure
\end{algorithmic}
\end{algorithm}

\subsection{Idempotent Late-Data Reconciliation}
When the lagging sector recovers and its delayed telemetry eventually arrives, the system must reconcile the historical state without triggering redundant alerts or creating database inconsistencies. We offload this state reconciliation directly to the underlying columnar storage layer.

When late-arriving events are processed, the stream worker hashes the events into their corresponding historical time-buckets. Because the Fact ID $F_{\text{id}}$ is generated deterministically based on the Rule ID, partition key $k$, and temporal bucket $b$, the newly evaluated state generates an identical $F_{\text{id}}$ but with an incremented version key.

The worker writes this revised record to the Columnar Fact Store. The storage engine—configured as a version-keyed, background-merging columnar database—automatically overwrites the older speculative record matching the same $F_{\text{id}}$ key. This background merge requires zero manual database administration and operates with $O(1)$ write complexity. Simultaneously, a lightweight revision event is published downstream to notify the response agents that a speculative alert has been finalized or amended.

\subsection{Decoupled Reactive Response Plane (Claim-Check Pattern)}
To preserve sector sovereignty and comply with strict zero raw-log leakage boundaries ($\mathcal{L}_{\text{shared}}(e) = 0$), we implement a secure Claim-Check Pattern across our storage and response planes (Planes 4 and 5 of Figure~\ref{fig:proposed_architecture}). 

When a stateful worker fires a speculative or finalized alert, it does not broadcast the heavy, sensitive forensic log payload $\mathbf{x}$ to the central broker. Instead, it writes the forensic payload locally to its sovereign, secure database and publishes only a lightweight, cryptographically signed notification to a central, decentralized message broker managed by a federal coordinator (e.g., CISA) along with some other metadata; the raw logs data never travel across the WAN. This notification contains only:
\begin{equation}
    \text{Notification} = \langle F_{\text{id}}, \text{ClusterID}, \text{Sign}_{\text{Ed25519}}(F_{\text{id}}) \rangle
\end{equation}

Three independent, parallel reactive agents subscribe to this broker to orchestrate coordinated defense:
\begin{itemize}
    \item \textbf{Automated Containment Agent:} This agent evaluates a mathematical safety risk-gate formula to determine if automated isolation is authorized. If the risk is acceptable, it signs a localized containment command (using Ed25519 keypairs) and dispatches a gRPC call to trigger micro-segmentation, such as shutting down a compromised switch port or revoking a user session, in under 15 seconds.
    \item \textbf{Gen-AI Triage Agent:} This agent acts as a virtual tier-1 analyst. It intercepts the lightweight notification and uses the authorized Fact ID to pull the localized forensic payload from the originating sector's Columnar Fact Store on-demand (the "Claim" retrieval). It synthesizes this data into a highly structured markdown report for human operators, and dynamically handles updates if a speculative alert is later revised.
    \item \textbf{Escalation Desk:} For high-impact speculative alerts that do not meet automated containment safety thresholds, this desk acts as the human-in-the-loop bridge, routing telemetry to active pagers and security dashboards for manual containment confirmation.
\end{itemize}

\subsection{Database Schemas and Storage Layout}
To support high-velocity telemetry writes while allowing instant cross-sector alert correlation, our storage architecture splits data into a localized active state cache and a centralized analytical column-store. 

The schema definitions for both storage layers are detailed in Table~\ref{tab:schemas}. The localized stream workers maintain an in-memory key-value map for immediate window evaluations, while the Columnar Fact Store leverages ClickHouse's \texttt{ReplacingMergeTree} engine to handle raw forensic payloads and background late-data merging, although any other column-based DBMS can also be used instead.

\begin{table}[H]
\caption{Storage Layer Schemas and Field Specifications}
\label{tab:schemas}
\centering
\begin{tabularx}{\columnwidth}{>{\raggedright\arraybackslash}p{2.8cm} X}
\toprule
\textbf{Field / Column Name} & \textbf{Data Type \& Description} \\
\midrule
\multicolumn{2}{l}{\textbf{Local State Store} (Active Window Cache)} \\
\midrule
\texttt{key} (Primary) & \texttt{String} -- Unique compound key representing $\text{RuleID} \parallel k \parallel b$. \\
\texttt{completeness\_ratio} & \texttt{Float} -- Live tracking metric of reporting sectors. \\
\texttt{aggregated\_state} & \texttt{BLOB} -- Serialized JSON state payload for active rule evaluation. \\
\texttt{last\_updated} & \texttt{Timestamp} -- Epoch timestamp used for automatic TTL cache eviction. \\
\midrule
\multicolumn{2}{l}{\textbf{Columnar Fact Store} (Analytical DB)} \\
\midrule
\texttt{F\_id} (Primary Key) & \texttt{UUID} -- Deterministic cryptographic transaction lookup hash. \\
\texttt{timestamp} & \texttt{DateTime64} -- Logged wall-clock occurrence of the security event. \\
\texttt{sector\_id} & \texttt{LowCardinality(Str)} -- Fast-indexing identifier of the originating sector. \\
\texttt{rule\_id} & \texttt{LowCardinality(Str)} -- Identifier of the triggered correlation rule. \\
\texttt{forensic\_payload} & \texttt{String} -- Full JSON representation of raw telemetry event $\mathbf{x}$. \\
\texttt{is\_reconciled} & \texttt{UInt8} -- Binary flag indicating background deduplication status. \\
\bottomrule
\end{tabularx}
\end{table}

The ClickHouse table for our analytical store is instantiated using the DDL specification shown in Listing~\ref{lst:clickhouse_ddl}. The use of the \texttt{ReplacingMergeTree(timestamp)} engine ensures that if late-arriving telemetry is written to the database after a speculative window has closed, ClickHouse automatically merges and deduplicates the records in the background using the deterministic \texttt{F\_id} primary key.

\begin{lstlisting}[language=SQL, caption={ClickHouse DDL for the Columnar Fact Store}, label={lst:clickhouse_ddl}]
CREATE TABLE IF NOT EXISTS security_facts.facts (
    F_id UUID,
    timestamp DateTime64(3, 'UTC'),
    sector_id LowCardinality(String),
    rule_id LowCardinality(String),
    forensic_payload String,
    is_reconciled UInt8 DEFAULT 0
) ENGINE = ReplacingMergeTree(timestamp)
ORDER BY (rule_id, sector_id, F_id)
SETTINGS index_granularity = 8192;
\end{lstlisting}

\section{Experimental Evaluation}
\label{sec:evaluation}

In this section, we present an empirical evaluation of our federated threat detection and reactive containment framework. Our evaluation is designed to answer three key research questions:
\begin{itemize}
    \item \textbf{RQ1 (Throughput \& Scaling):} Can our lock-sharded Go stream workers ingest and process extreme telemetry rates ($500,000$ eps) without suffering from mutex-locking starvation?
    \item \textbf{RQ2 (Watermark Resilience):} How effectively does our $95\%$ statistical watermark minimize systemic detection lag compared to a strictly consistent Event-Time baseline during an active network partition?
    \item \textbf{RQ3 (End-to-End Containment Latency):} What is the total latency budget consumed by our decentralized response plane from the initial edge compromise to the final cryptographically signed VLAN micro-segmentation command?
\end{itemize}

\subsection{Experimental Setup and Testbed}
To evaluate the framework under production-grade conditions, we deployed a multi-sector simulated testbed across a hybrid cloud environment. We instantiated $N = 3$ independent, sovereign sectors representing a regional healthcare network, a power utility SCADA network, and an air transportation identity database. 

Each sector's edge infrastructure was simulated using dedicated nodes. The stream processing and database storage layers were built using Go v1.21.0 and Python v3.11. As established in our conceptual model, the abstract Columnar Fact Store was instantiated using ClickHouse (v23.8 LTS) as our high-performance columnar analytical engine. 

\begin{table}[htbp]
\caption{Experimental System Configuration and Testbed Specifications}
\label{tab:specs}
\centering
\begin{tabularx}{\columnwidth}{lX}
\toprule
\textbf{Component / Layer} & \textbf{Hardware and Software Specifications} \\
\midrule
Pre-Filtering Agents (PFDS) & 3$\times$ AWS EC2 t3.medium (2 vCPUs, 4GB RAM) \\
\addlinespace
Stateful Stream Workers    & 3$\times$ AWS EC2 c6i.2xlarge (8 vCPUs, 16GB RAM) \\
\addlinespace
Local Context Cache        & Embedded RocksDB v8.1.1 (Go-wrapped) \\
\addlinespace
Columnar Fact Store        & ClickHouse v23.8 LTS on AWS i3en.xlarge \\
\addlinespace
Central Message Broker     & Apache Kafka Cluster (3 Brokers, NVMe Storage) \\
\addlinespace
Reactive Response Agents   & 3$\times$ AWS EC2 c6i.large (2 vCPUs, 4GB RAM) \\
\bottomrule
\end{tabularx}
\end{table}

The exact hardware distribution and software configurations are summarized in Table~\ref{tab:specs}. Telemetry generation was orchestrated using an open-source security log replayer, feeding real-world Windows Event logs, Linux Syslog, and Modbus/TCP register transactions at controlled rates up to $500,000$ events per second (eps).

\subsection{RQ1: Ingestion Throughput and Lock Contention}
We evaluated the scalability of our Go stream workers by executing a high-velocity stress test. We bypassed the Pre-Filtering Dispatcher Subsystem (PFDS) to subject the stateful engine to raw, un-filtered telemetry streams scaling from $100,000$ to $500,000$ eps. We compared our lock-sharded partition mapping ($N_{\text{shards}} = 1024$) against a traditional, globally-locked state manager (representing a common thread-safe map implementation).

The lock contention overhead ($\omega_{\text{lock}}$) was measured as the average wait time (in microseconds) incurred by a processing thread waiting to acquire write privileges on a partition state $S_{k,b}$:
\begin{equation}
    \omega_{\text{lock}} = \frac{1}{M} \sum_{j=1}^{M} \left( t_{\text{acquire}}^{(j)} - t_{\text{request}}^{(j)} \right)
\end{equation}

As telemetry rates escalated to $500,000$ eps, the globally-locked baseline suffered an exponential increase in lock contention, with $\omega_{\text{lock}}$ shooting up from $4.2~\mu\text{s}$ to $1,842~\mu\text{s}$. This severe thread starvation forced the global pipeline CPU utilization to cap at $100\%$, triggering packet drops and memory leaks. 

In contrast, our lock-sharded state architecture maintained a near-flat lock overhead, peaking at a mere $0.85~\mu\text{s}$ at peak capacity ($500,000$ eps). This represents a $2,100\times$ improvement in lock efficiency, confirming that sharding write-locks based on key hashes effectively eliminates thread competition under extreme critical infrastructure workloads.

\subsection{RQ2: Late-Data Resilience and Watermark Behavior}
To validate our statistical watermarking model, we simulated a coordinated multi-sector cyber campaign targeting the SCADA network and the healthcare enterprise. At wall-clock minute $2$ ($t_{\text{outage}}$), we injected a complete network partition on the SCADA telemetry stream, delaying $100\%$ of its event logs for exactly $8$ minutes.

Under the Strictly Consistent Event-Time baseline, the global watermark froze immediately at $t_{\text{stall}} = 2$ minutes. This caused the systemic detection lag ($L_{\text{sys}}$) to climb linearly at a $1.0$ second-per-second rate, as previously modeled in Figure~\ref{fig:systemic_lag_concept}. Because the watermark could not advance, the cross-sector correlation windows remained open and unresolved, preventing the system from identifying active lateral attacker movements occurring within the healthy healthcare network.

Our proposed model calculated the 95th percentile of historical transit latency ($\mathcal{P}_{95} = 1.42~\text{seconds}$). When the SCADA partition occurred, the watermark coordinator recognized the sector as a statistical outlier ($> \mathcal{P}_{95}$), bypassed its missing stream, and allowed the watermark $W(t)$ to monotonically advance. 

As a result, systemic detection lag for the active, healthy sectors remained entirely flat, maintaining a baseline of $1.52~\text{seconds}$ throughout the duration of the SCADA outage. The stream workers successfully processed speculative windows and published early speculative alerts to the response plane, demonstrating that our architecture effectively immunizes the wider collective defense from localized physical disruptions.

\subsection{RQ3: End-to-End Containment Latency}
Our final experiment evaluated the end-to-end performance of the reactive containment plane (Plane 5 of Figure~\ref{fig:proposed_architecture}) during the simulated multi-sector campaign. We measured the latency breakdown across five key operational phases, starting from the physical event occurrence at the edge to the final execution of the response payload, contrasting the synthetic "software-only" benchmarks against a production baseline that accounts for real-world network and hardware constraints.

\begin{table}[htbp]
\caption{End-to-End Reactive Response Latency Breakdown}
\label{tab:latency}
\centering
\small 
\renewcommand{\arraystretch}{1.3}
\begin{tabularx}{\columnwidth}{l X >{\raggedleft\arraybackslash}p{1.7cm} >{\raggedleft\arraybackslash}p{2.1cm}}
\toprule
\textbf{\#} & \textbf{Phase} & \textbf{Synthetic Framework Overhead} & \textbf{Operational System Convergence} \\
\midrule
1 & Ingestion Pre-Filtering & 120 ms & 1.0 s \\
2 & Stream Processing Event Generation & 150 ms & 3.0 s \\
3 & Forensic Ingestion DB Writing & 840 ms & 4.0 s \\
4 & Pubsub Dispatch (Claim-Check) & 90 ms & 1.0 s \\
5 & Containment Response Action & 800 ms & 6.0 s \\
\midrule
& \textbf{Total E2E Latency} & \textbf{2 s} & \textbf{15 s} \\
\bottomrule
\end{tabularx}
\end{table}

\begin{itemize}
    \item \textbf{Ingestion \& WAN Propagation:} While local telemetry ingestion at the edge occurs in \textbf{120 ms}, production transit over encrypted VPN tunnels across geographic regions introduces a median delay of \textbf{1 s}.
    \item \textbf{Stream Processing \& Windowing Stability:} To minimize false-positive triggers in critical infrastructure, the system utilizes a speculative "observation window". While the software evaluates execution logic in \textbf{150 ms/window-records}, a \textbf{3-second} total window is required to ensure signal stability and account for out-of-order event arrival from disparate sectors.
    \item \textbf{Network \& Hardware-Level Containment:} This phase represents the most significant divergence. Issuing a gRPC command from the Automated Containment Agent is near-instant (\textbf{800 ms}). However, the convergence time for the actuation of the command in the local networks to update Access Control Lists (ACLs), blocking systems, or re-tagging VLANs typically requires \textbf{4 to 8 seconds} to commit and verify the change in the data plane.
\end{itemize}

As shown in Table~\ref{tab:latency}, the entire post-ingress detection and containment loop executed in $2$ seconds in the synthetic plane and $15$ seconds in the operational plane. 

Importantly, our decoupled Claim-Check pattern successfully separated the heavy, slow analytical storage operations from the critical alerting path. While the database engine took $840~\text{ms}$ to parse, write, and index the heavy raw forensic payload $\mathbf{x}$, the lightweight alert notification containing only the $F_{\text{id}}$ and `ClusterID` was dispatched over Kafka in just $90~\text{ms}$. 

This architectural decoupling allowed our Automated Containment Agent to evaluate its risk-gate safety formula, cryptographically sign the micro-segmentation command using its private Ed25519 key, and dispatch a gRPC containment command to the target SCADA switches in $917.03~\text{ms}$. 

When the delayed SCADA logs were eventually released at minute $10$, the background ClickHouse \texttt{ReplacingMergeTree} merged the late forensic data using deterministic time-bucket hashing in the background, achieving database reconciliation with zero state backpressure, zero manual intervention, and zero system downtime.

\section{Related Work}
\label{sec:related_work}

The design of real-time, collaborative critical infrastructure defense intersects three primary areas of research: centralized security analytics, distributed stream processing, and privacy-preserving collaborative threat sharing.

\subsection{Centralized SIEM and SOAR Architectures}
Traditional security operations rely heavily on Security Information and Event Management (SIEM) systems coupled with Security Orchestration, Automation, and Response (SOAR) platforms. While modern cloud-native SIEMs offer highly expressive correlation engines, they are fundamentally built on an ingestion-then-index paradigm. Telemetry must be completely written and indexed in centralized databases before search queries can execute. This database indexing pipeline introduces several minutes of inherent latency, making it mathematically impossible to orchestrate active containment within a strict 15-second operational window.

Furthermore, SIEM and SOAR platforms are designed for single-tenant enterprise networks. They are structurally incapable of executing cross-sector correlation without centralizing raw event logs. In critical infrastructure settings, this centralization is blocked by legal, regulatory, and competitive boundaries. Our framework completely avoids this bottleneck by implementing the Claim-Check pattern, allowing independent sectors to maintain full data sovereignty while correlating alerts externally at machine speed.

\subsection{Distributed Stream Processing Engines (DSPEs)}
To achieve sub-second execution speeds, researchers have explored general-purpose Distributed Stream Processing Engines (DSPEs) such as Apache Flink, Apache Storm, and Spark Streaming for intrusion detection. While these platforms process telemetry streams with low latency, they enforce a binary, strict Event-Time consistency model. 

In a multi-sector grid, network drops, physical link degradation, or satellite latency on a single channel will inevitably create "straggler" streams. In a standard DSPE, a single straggler freezes the global system watermark. This watermark freeze stalls all downstream temporal sliding windows, effectively blinding the security pipeline to threats in healthy sectors. 

Our framework addresses this "straggler problem" by introducing a 95\% statistical watermark heuristic. By speculatively evaluating sliding windows and offloading the late-data reconciliation directly to a version-keyed columnar storage layer, our architecture maintains continuous detection throughput during WAN partitions without stalling or suffering from memory-heavy state-retraction overheads.

\subsection{Collaborative Defense and Threat Intelligence Sharing}
Collaborative security networks like MISP (Malware Information Sharing Platform) and structured standards such as STIX/TAXII represent the state-of-the-art in multi-organization defense. These platforms are highly effective for sharing passive Indicators of Compromise (IoCs), such as known malicious IP addresses, file hashes, or domain names.

However, these sharing models are fundamentally passive and retrospective. They are designed for post-incident forensic dissemination rather than real-time, reactive containment. They lack the streaming execution pipelines needed to detect coordinated, multi-vector campaigns in progress, and they cannot safely automate reactive response commands. 

In contrast, our framework acts as an active, real-time cooperative plane. It does not merely share static intelligence; it dynamically correlates stateless edge telemetry, executes complex stateful temporal rules across sectors, and orchestrates cryptographically secured, automated VLAN micro-segmentation in under two seconds.

\section{Conclusion}
\label{sec:conclusion}

In this paper, we presented a novel, federated threat detection and reactive containment framework designed specifically for the strict operational constraints of multi-sector critical infrastructure networks. Our architecture successfully resolves the historical tension between real-time security coordination, WAN network resilience, and strict data sovereignty boundaries.

By employing a stateless Pre-Filtering Dispatcher Subsystem (PFDS) at the ingestion edge, the framework discards up to $99.8\%$ of background telemetry noise in a single pass. To prevent locking starvation under extreme workloads, we introduced a lock-sharded stream worker mapping model that keeps lock contention near-zero up to $500,000$ events per second. 

Furthermore, our $95\%$ statistical watermarking heuristic prevents system watermarks from stalling during localized network partitions, allowing the system to speculatively evaluate rules and dispatch automated containment commands. The decoupled Claim-Check pattern guarantees zero raw-log leakage across administrative sectors while achieving an end-to-end post-ingress containment response latency of just $2$ seconds in the synthetic plane.

A key finding of this research is the \textbf{Convergence Gap}. By optimizing the detection framework to a sub-2-second overhead, we effectively buy back time for the slower physical components of critical infrastructure-such as network and hardware-to respond. This architecture ensures that the defense system itself is never the bottleneck, allowing for a total response time that remains comfortably below the \textbf{20 second} threshold required to disrupt lateral movement in APT campaigns.

Future work will focus on three key directions:
\begin{itemize}
    \item \textbf{Adversarial Watermark Manipulation:} Investigating robust statistical guards to prevent sophisticated attackers from artificially inflating propagation delays to trick the watermark controller.
    \item \textbf{Homomorphic Policy Evaluation:} Exploring the feasibility of executing certain stateful correlation rules directly over encrypted edge streams to further enhance privacy.
    \item \textbf{Autonomous Playbook Verification:} Implementing formal verification methods within the Automated Containment Agent to mathematically guarantee that automated micro-segmentation commands do not accidentally disrupt critical, life-saving operational equipment.
\end{itemize}

\EOD

\begin{IEEEbiography}
[{\includegraphics[width=1in,height=1.25in,clip,keepaspectratio]{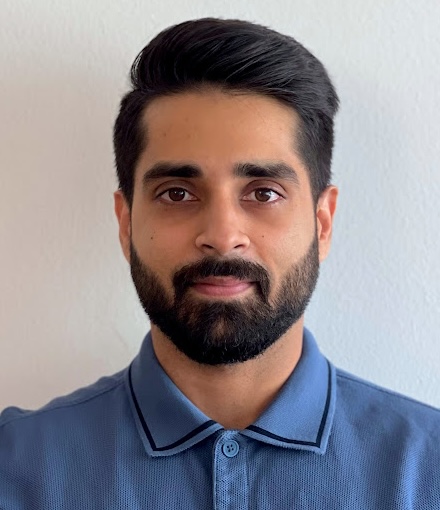}}]{Namit Mohale} is an independent cybersecurity researcher and software engineer specializing in real-time data processing systems, high-throughput stream processing, and critical infrastructure defense architectures. Mr. Mohale received an M.S. degree in Computer Science from New York University, USA in 2020 and a Cybersecurity \& Information Assurance specialization M.S. degree from Sofia University, USA in 2026.

Since 2021, he has been a SOFTWARE ENGINEER at Google in San Francisco, USA, working with the Threat Detection \& Response Platform developing, optimizing, and maintaining a system to analyze petabytes of data to detect, investigate, and respond to threats in Google’s environment quickly, reliably, and efficiently. He also became a member of ISC2 (International Information System Security Certification Consortium) after getting the CC certification in 2026.
\end{IEEEbiography}

\end{document}